\documentclass[twocolumn,superscriptaddress,amsmath,amssymb,aps,pra]{revtex4-1}

\usepackage{graphicx}
\usepackage{bm}

\usepackage{color}

\usepackage{graphicx}
\usepackage{comment}
\usepackage{color}
\usepackage[colorlinks,urlcolor=blue,citecolor=blue,linkcolor=blue]{hyperref}
\usepackage{cleveref}
\usepackage{physics}
\usepackage{mathtools}
\usepackage{mathrsfs}
\usepackage{booktabs}

\newcommand{\kv}{\vectorbold{k}}
\newcommand{\qv}{\vectorbold{q}}
\newcommand{\med}{0}
\newcommand{\ch}{\hat{c}}
\newcommand{\chd}{\hat{c}^\dagger}
\newcommand{\fh}{\hat{f}}
\newcommand{\fhd}{\hat{f}^\dagger}
\renewcommand{\dh}{\hat{d}}
\newcommand{\dhd}{\hat{d}^\dagger}
\newcommand{\rhoh}{\hat{\rho}}
\newcommand{\GT}{\Tilde{G}}
\newcommand{\omegab}{\bar{\omega}}
\newcommand{\Cb}{\bar{C}}
\newcommand{\Arg}{\text{Arg}}
\newcommand{\GM}{\mathcal{G}}

\usepackage{tikz}
\usetikzlibrary{calc,decorations.markings}

\begin{document}

\title{Signatures of the orthogonality catastrophe in a coherently driven impurity}

\author{Haydn S. Adlong}
\affiliation{School of Physics and Astronomy, Monash University, Victoria 3800, Australia}
\affiliation{ARC Centre of Excellence in Future Low-Energy Electronics Technologies, Monash University, Victoria 3800, Australia}

\author{Weizhe Edward Liu}
\affiliation{School of Physics and Astronomy, Monash University, Victoria 3800, Australia}
\affiliation{ARC Centre of Excellence in Future Low-Energy Electronics Technologies, Monash University, Victoria 3800, Australia}

\author{Lincoln D. Turner}
\affiliation{School of Physics and Astronomy, Monash University, Victoria 3800, Australia}

\author{Meera M. Parish}
\affiliation{School of Physics and Astronomy, Monash University, Victoria 3800, Australia}
\affiliation{ARC Centre of Excellence in Future Low-Energy Electronics Technologies, Monash University, Victoria 3800, Australia}

\author{Jesper Levinsen}
\affiliation{School of Physics and Astronomy, Monash University, Victoria 3800, Australia}
\affiliation{ARC Centre of Excellence in Future Low-Energy Electronics Technologies, Monash University, Victoria 3800, Australia}

\begin{abstract}
We consider a fixed impurity immersed in a Fermi gas at finite temperature. We take the impurity to have two internal spin states, where the $\uparrow$ state is assumed to interact with the medium such that it exhibits the orthogonality catastrophe, while the $\downarrow$ state is a bare noninteracting particle. Introducing a Rabi coupling 
between the impurity states therefore allows us to investigate the coupling between a discrete spectral peak and the Fermi-edge singularity, i.e., between states with and without a quasiparticle residue.
Combining an exact treatment of the uncoupled impurity Green's functions 
with a variational approach to treat the Rabi driven dynamics, we find that
the system features 
Rabi oscillations whose frequency scales as a non-trivial power of the Rabi drive at low temperatures.
This reflects the power law of the Fermi-edge singularity
and, importantly, this behavior is qualitatively different from the case
of a mobile impurity quasiparticle where the scaling is linear. We
therefore argue that the scaling law 
serves as an
experimentally implementable probe of the orthogonality catastrophe.
We additionally %
simulate rf spectroscopy beyond linear response, finding a remarkable agreement
with an experiment using heavy impurities [Kohstall \textit{et al.}, Nature \textbf{485}, 615 (2012)], thus demonstrating the power of our approach.
\end{abstract}

\date{\today}

\maketitle

\section{Introduction}

Quasiparticles have been a cornerstone of quantum many-body theory since Landau first introduced them in the context of ${}^3$He at low temperatures~\cite{Pines}. %
However, understanding systems beyond the quasiparticle paradigm remains challenging.
A notable example of a problem that defies a quasiparticle description is the response of a free Fermi gas to a static potential. This problem directly arises in the context of x-ray spectra of metals~\cite{nozieres1969singularities,mahan2013many} and 
is connected to the physics of the Kondo effect~\cite{Anderson1970,latta2011quantum,VonDelft2011}.
The response is governed by the orthogonality catastrophe --- i.e., the orthogonality between the many-body ground-state wave functions of the Fermi gas with and without the potential~\cite{Anderson1967} --- which translates into a Fermi-edge singularity in the spectral response~\cite{mahan2013many}. 

\begin{figure}
    \centering
    \includegraphics[width=0.85\linewidth]{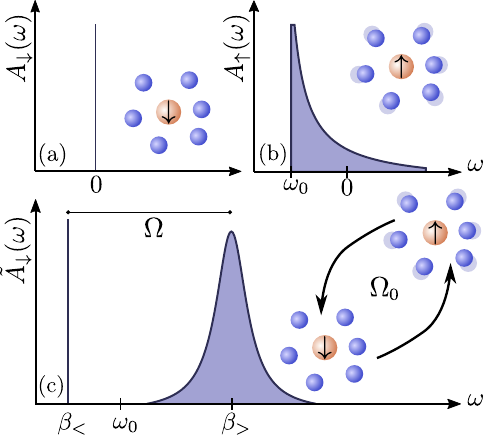}
    \caption{Schematic illustration of the effects of Rabi driving on the spectral function. In the top row, the impurity spin states are decoupled. Consequently, the $\downarrow$ spectral function $A_\downarrow(\omega)$ is a delta function centred at zero energy (a), while the $\uparrow$ spectral function $A_\uparrow(\omega)$ exhibits a Fermi-edge singularity at $\omega_0$ (b). In panel (c), the $\downarrow$ spectral function is evaluated in the presence of Rabi coupling to the $\uparrow$ state. The coupled spectrum $\tilde{A}_\downarrow(\omega)$ reveals the emergence of two 
    quasiparticle peaks at $\beta_<$ and $\beta_>$.
    The frequency difference $\Omega$ between the peaks encodes details of the Fermi-edge singularity.}
    \label{fig:Schematic}
\end{figure}

Ultracold atomic gases provide a particularly versatile platform for probing and investigating the orthogonality catastrophe~\cite{Goold2011,Knap2012,Dora2013,Sindona2013,Schiro2014,Campbell2014,Deng2015,Schmidt2018,Mistakidis2019,You2019,Fogarty2020}
due to the unparalleled control over parameters such as the atom-atom interactions~\cite{bloch2012quantum}.
Here, a static potential in a Fermi gas can be created by pinning an impurity atom with an optical field  
and using a particular impurity spin state (denoted $\uparrow$) that is strongly interacting with the surrounding fermions.
Such a setup %
represents the infinite-mass limit %
of the scenario of a mobile impurity coupled to a quantum gas~\cite{Massignan2014review}, which has been successfully realized for both Fermi~\cite{Schirotzek2009,Nascimbene2009,Kohstall2012,Koschorreck2012,Zhang2012,Wenz2013,Cetina2015,Ong2015,Cetina2016,Scazza2017,Yan2019,Oppong2019,Ness2020} and Bose gases~\cite{Catani2012,Hu2016,Jorgensen2016,Camargo2018,Yan2019u,Skou2021}.
A common protocol to probe the coherence of a mobile impurity quasiparticle 
is Rabi oscillations between the $\uparrow$ state and an auxiliary, noninteracting $\downarrow$ state~\cite{Kohstall2012,Scazza2017,Oppong2019}. In particular, the ratio of the frequency of the Rabi oscillations $\Omega$ to the Rabi drive $\Omega_0$ yields the quasiparticle residue via $Z \approx (\Omega/\Omega_0)^2$~\cite{Kohstall2012}. %
In the case of a fixed impurity, such Rabi oscillations couple the Fermi-edge singularity of the $\uparrow$ spectrum with the Dirac-delta peak
of the noninteracting $\downarrow$ spectrum (Fig.~\ref{fig:Schematic}). %
Since the orthogonality catastrophe features a vanishing quasiparticle residue ($Z \to 0$), %
the Rabi driven fixed impurity must either cease to display coherent oscillations or it must undergo fundamental changes from the mobile impurity case such that $Z \not\approx  (\Omega/\Omega_0)^2$.

In this paper, we use a combination of exact and %
variational approaches to explore the Rabi driven dynamics %
of an infinitely heavy impurity in a Fermi gas. 
In the absence of Rabi coupling ($\Omega_0=0$), the $\uparrow$ and $\downarrow$ subsystems decouple and the problem can be solved exactly~\cite{Goold2011,Knap2012} --- in the non-trivial, interacting $\uparrow$ subsystem, the exact solution is found using a functional determinant approach~\cite{Levitov1993}. On the other hand, the inclusion of Rabi coupling results in a system which is no longer exactly solvable. 
However, within a time-dependent variational approach~\cite{Parish2016,Liu2019}, we find that the dynamics can be related to the spin-decoupled $\uparrow$ and $\downarrow$ Green's functions, which enables us to incorporate their exact solutions. 
Within this approximation, we find that the Rabi coupled system exhibits coherent oscillations, which are linked to the remarkable emergence of quasiparticles in the coupled $\downarrow$ impurity spectrum (see Fig.~\ref{fig:Schematic}). Furthermore, properties of the orthogonality catastrophe are encoded in these emergent quasiparticles. Specifically, we find that that the energy difference between the quasiparticles --- 
which yields the Rabi oscillation frequency %
$\Omega$ --- scales as a power law of the drive frequency $\Omega_0$, where the power law exponent is directly related to that governing the Fermi-edge singularity. 
This is in contrast to the linear scaling expected for a mobile impurity, and we therefore argue that this scaling serves as a signature of the orthogonality catastrophe.

To further demonstrate the utility of our theory, we investigate the regime of weak Rabi drive with respect to the Fermi energy, where our theory is expected to be accurate. In particular, we obtain exact analytic relations for the short-time dynamics that are valid for both Bose and Fermi mediums, similar to Ref.~\cite{Parish2016} for the case of Ramsey interferometry.
Finally, we provide the first simulation of radio-frequency (rf) spectroscopy for an impurity in a Fermi gas beyond linear response and we find excellent agreement with an experiment that used heavy $^{40}$K impurities in a $^{6}$Li gas~\cite{Kohstall2012}. 

This paper is organized as follows. In Sec.~\ref{Sect:Theory}, we introduce our model %
and derive an expression for 
the variational impurity Rabi oscillations in terms of the exact spin-decoupled Green's functions. Here, we also include %
the exact $\uparrow$ Green's function, as calculated using a %
functional determinant approach. In Sec.~\ref{Sect:Applications} we 
calculate the short-time dynamics of Rabi oscillations, and we simulate rf spectroscopy beyond linear response. In Sec.~\ref{Sect:OrthogonalityCatastrophe}, we answer the question of what can be learnt through coupling a discrete spectral peak to the Fermi-edge singularity. In particular, we derive an analytical model of the effect of the orthogonality catastrophe on Rabi oscillations and compare this against our variational Rabi oscillations. We conclude in Sec.~\ref{Sect:Conclusion}. Details on the spin-coupled Green's function, the short-time dynamics, and the analytical model are contained in Appendix~\ref{App:DetailsGreenFunc}, \ref{Sect:PertRabi} and \ref{Sect:AnalyticalModel}, respectively.

\section{Theoretical framework} \label{Sect:Theory}

\subsection{Model}
We model Rabi oscillations between two spin states ($\sigma = \uparrow, \downarrow$) of a fixed impurity in a Fermi gas at temperature $T$ using a Hamiltonian consisting of three terms:
\begin{align} \label{Eq:GenericHam}
    \hat{H} = \hat{H}_{\med} + \hat{H}_{\rm int} %
    + \hat{H}_\Omega.
\end{align}
In what follows we will work in units in which $\hbar$, $k_B$ and the system volume %
are all set to one. To account for the temperature of the system, we employ a grand-canonical formulation of the Fermi gas
\begin{align}
    \hat{H}_{\med} = \sum_{\kv} (\epsilon_{\kv} - \mu) \fhd_{\kv} \fh_{\kv},
\end{align}
where $\fhd_{\kv}$ creates a majority fermion with momentum $\kv$ and mass $m$, $\mu$ is the corresponding chemical potential and $\epsilon_{\kv} = |\kv|^2/2m=k^2/2m$ the dispersion. The Fermi momentum $k_F$ %
parameterizes the medium with Fermi energy $E_F \equiv k_F^2/2m$, Fermi temperature $T_F \equiv E_F$ and Fermi time $\tau_F \equiv 1/E_F$.

The term $\hat{H}_{\rm int}$ captures the interactions between the $\uparrow$ impurity state and the Fermi medium. We use a two-channel Hamiltonian~\cite{Timmermanns1999}
\begin{align} \label{Eq:Hup2Channel}
    \hat{H}_{\rm int} = \nu \dhd \dh + g \sum_{\kv}^\Lambda \left( \dhd \ch_\uparrow \fh_{\kv} + \fhd_{\kv}  \chd_\uparrow \dh   \right),
\end{align}
where $\ch_\sigma$ is the $\sigma$ impurity annihilation operator. In Eq.~\eqref{Eq:Hup2Channel} the interactions between the $\ch_\uparrow$ and $\fh$ particles are mediated via the exchange of a closed-channel dimer $\dh$, where we have a coupling constant $g$ (up to the ultra-violet cutoff $\Lambda$) and closed-channel detuning $\nu$. These bare parameters can be related to physical parameters --- namely, the $s$-wave scattering length $a$ and effective range $R^*$ --- through the process of renormalization:
\begin{align}
    \frac{m}{2\pi a } = -\frac{\nu}{g^2} + \sum_{\kv}^{\Lambda} \frac{1}{\epsilon_{\kv}}, \qquad R^* = \frac{\pi}{m^2 g^2}.
\end{align}
For positive scattering lengths ($a>0$) there exists an impurity-fermion bound state with energy
\begin{align} \label{Eq:BoundState}
    E_b = - \frac{\left( \sqrt{1+4R^*/a}-1 \right)^2}{8m R^*{}^2}.
\end{align}
On the other hand, %
the $\downarrow$ impurity state $\ch_\downarrow$ is assumed to have negligible interactions with the Fermi gas.

Finally, $\hat{H}_\Omega$ is the Hamiltonian of the continuous semi-classical radiation field 
that drives the Rabi oscillations. Working in the rotating-wave approximation, $\hat{H}_\Omega$ couples the impurity spin states according to
\begin{align} \label{Eq:RotWave}
    \hat{H}_\Omega = \frac{\Omega_0}{2} ( \chd_\uparrow \ch_\downarrow  + \chd_\downarrow \ch_\uparrow) + \Delta \chd_\downarrow \ch_\downarrow,
\end{align}
where $\Omega_0$ is the bare Rabi coupling and $\Delta$ is the detuning, which is measured from the bare $\uparrow$-$\downarrow$ transition.

\subsection{Variational Rabi oscillations} \label{Sect:VarRabiOsc}
Assuming the Rabi coupling is switched on at $t=0$, %
the impurity population in state $\sigma$ at times $t\geq 0$ is given by
\begin{align} \label{Eq:RabiFormula}
    \mathcal{N}_\sigma (t) = \Tr [ \rhoh_0 \ch (t) \chd_\sigma \ch_\sigma \chd (t)].
\end{align}
Here, $\ch(t)$ is the impurity operator in the Heisenberg picture, $\hat \rho_0=\exponential(-\beta \hat
H_0)/\Trace[\exponential (-\beta \hat H_0)]$ (with $\beta\equiv1/T$) is the medium density matrix and the trace is taken over all realizations of the medium in the absence of the impurity. We will consider the impurity initialized in the $\downarrow$ state, i.e., $\ch(t=0) = \ch_\downarrow$, such that we have $\mathcal{N}_\downarrow (0) =1$ and $\mathcal{N}_\uparrow (0) = 0$. Note that we always have $\mathcal{N}_\downarrow (t) + \mathcal{N}_\uparrow (t) = 1$.

In order to approximate the impurity Rabi oscillations, we calculate the dynamics in a truncated basis in which the impurity operator admits the form
\begin{align} \label{Eq:chAnsatz}
    \ch(t) &= \alpha_0^\downarrow(t) \ch_\downarrow + \alpha_0^\uparrow(t) \ch_\uparrow + \sum_{\kv} \alpha_{\kv}^\uparrow(t) \fhd_{\kv} d \nonumber \\
    &{} \hspace{5em}+ \sum_{\kv \qv} \alpha_{\kv \qv}^\uparrow(t) \fhd_{\qv} \fh_{\kv} \ch_{\uparrow} + \dots,
\end{align}
where the expansion coefficients $\{\alpha^\sigma_j\}$ are complex functions of time, and the ellipsis includes all higher order particle-hole excitations describing the $\uparrow$ impurity interactions with the Fermi gas. Here, the operator basis $\{ \hat{O}_j \} = \{\ch_\downarrow , \ch_\uparrow, \fhd_{\qv} \fh_{\kv} \ch_{\uparrow}, \dots \}$ satisfies the orthogonality condition $\Tr[\rhoh_{\med}\hat{O}_j \hat{O}^\dagger_{j'} ] =0$ for $j\neq j'$, since the trace is over medium-only states.
Inserting the truncated operator \eqref{Eq:chAnsatz} %
in Eq.~\eqref{Eq:RabiFormula} then yields the \textit{variational Rabi oscillations}
\begin{align} \label{Eq:RabiVar}
    \mathcal{N}_\downarrow (t) \simeq | \alpha^\downarrow_0(t)|^2.
\end{align}    

In principle, the expansion coefficients $\{\alpha^\sigma_j \}$ can be calculated from minimizing the `error quantity'~\cite{Parish2016,Liu2019}
\begin{align}
    \mathcal{E}(t) \equiv \Tr[\rhoh_0\hat\epsilon (t)\hat\epsilon^\dagger (t)],
\end{align}
where the operator $\hat\epsilon (t) \equiv i\partial_t \ch(t) - \comm{\ch(t)}{\Hat{H}}$. Such an approach was used in Ref.~\cite{Adlong2020} to accurately model the experimentally measured Rabi oscillations of mobile impurities~\cite{Scazza2017,Oppong2019}. However, as we will now show, the dynamics within the variational operator basis, Eq.~\eqref{Eq:chAnsatz}, can be directly expressed in terms of the decoupled Green's functions, which can be calculated exactly.

To proceed, we introduce the Rabi coupled, retarded, $\downarrow$ Green's function $\Tilde{\GM}_\downarrow(t)$, which is related to the variational coefficients via
\begin{align}
    \Tilde{\GM}_\downarrow(t) = -i \Theta(t)\Tr[\rhoh_0 \ch(t) \chd_\downarrow] =  -i \Theta(t) \alpha_0^\downarrow (t),
    \label{eq:Gdownt}
\end{align}
where $\Theta(t)$ is the Heaviside function, and where we have used the fact that $\ch(t=0) = \ch_\downarrow$. Comparing with Eq.~\eqref{Eq:RabiVar}, we thus have 
\begin{align} \label{Eq:RabiFromGreen}
 \mathcal{N}_\downarrow (t) & \simeq |\Tilde{\GM}_\downarrow(t)|^2.   
\end{align}

The $\downarrow$ Green's function is easiest to determine in the frequency domain, which is obtained via the Fourier transform:
\begin{align} \label{Eq:InvFourierTransform}
    \Tilde{G}_\downarrow (\omega) = \int dt \, e^{i \omega t} \Tilde{\GM}_\downarrow (t) . 
\end{align}
In this case, the total Rabi coupled Green's function can be approximated from its spin decoupled (i.e., $\Omega_0=0$) counterparts:
\begin{align}\label{Eq:GCoupledRelation}
    \Tilde{G}(\omega) \simeq \mqty( G_\uparrow^{-1}(\omega) & \Omega_0/2 \\ \Omega_0/2 & G_\downarrow^{-1}(\omega))^{-1}.
\end{align}
Here, $\Tilde{G}_\downarrow(\omega) \equiv \Tilde{G}_{(2,2)}(\omega)$ and $G_\sigma(\omega)$ is the Fourier transform of the time-dependent $\sigma$ impurity Green's function, 
\begin{align}
    \GM_\sigma(t) = -i \Theta(t)\left. \Tr[\rhoh_0 \ch_\sigma e^{i \hat{H}_0 t} e^{-i \hat{H} t} \chd_\sigma] \right|_{\Omega_0 = 0}  \, ,
\end{align}
which is calculated in the absence of Rabi coupling.  
In writing Eq.~\eqref{Eq:GCoupledRelation}, we again neglect excitations in the Fermi gas during %
spin flipping to the $\downarrow$ state. Consequently, Eq.~\eqref{Eq:GCoupledRelation} is exact within the variational operator basis, Eq.~\eqref{Eq:chAnsatz}, which we prove in Appendix~\ref{App:DetailsGreenFunc}.

The variational ansatz in Eq.~\eqref{Eq:chAnsatz} represents the minimal form which incorporates both the orthogonality catastrophe --- requiring an infinite number of low-energy excitations --- and the coupling to a bare impurity particle. Specifically, it neglects the possibility of the $\downarrow$ impurity coexisting with excitations of the medium and hence it only approximately solves the Heisenberg equation of motion. At the level of the coupled impurity Green's function, our approximation leads to a $\downarrow$ self energy $\frac{\Omega_0^2}4G_\uparrow(\omega)$, which
 scales as $E_F(\Omega_0/E_F)^2$ in the absence of energy scales related to the interaction, e.g., at unitarity. 
On the other hand, the contribution to the $\downarrow$ self energy arising from terms that include co-propagation of the $\downarrow$ impurity with excitations scales as $E_F(\Omega_0/E_F)^4$ in the limit of small $\Omega_0/E_F$~\footnote{Contributions to the $\downarrow$ self energy arising from terms that include co-propagation of the $\downarrow$ impurity with excitations of the Fermi sea %
require a total of four Rabi spin flips: $\downarrow$ to $\uparrow$ followed by the creation of excitations, then $\uparrow$ to $\downarrow$ and back again without changing the excitations, and finally the excitations are re-absorbed and the $\uparrow$ state is flipped to $\downarrow$.}. Therefore, while our approximation is not exact, we expect it to be valid as long as $\Omega_0/E_F\ll 1$. Furthermore, the approximation becomes exact at early times $t\ll \tau_F$, since the excitations in the medium require time to accumulate.

\subsection{Spin decoupled Green's functions} \label{Sect:SpinDecGreenFunc}
Equations~\eqref{Eq:RabiFromGreen} and \eqref{Eq:GCoupledRelation} show that the variational Rabi oscillations can be calculated from a knowledge of the spin decoupled Green's functions $G_\sigma$, which we now show can be calculated exactly. For the noninteracting $\downarrow$ impurity, its Green's function is simply the `bare' propagator
\begin{align} \label{Eq:GreenDownBare}
    G_\downarrow (\omega) = \frac{1}{\omega - \Delta + i0},
\end{align}
where $\Delta$ is the detuning from the bare $\uparrow$-$\downarrow$ transition [see Eq.~\eqref{Eq:RotWave}]. Note that this corresponds to a retarded Green's function and thus involves an infinitesimal shift in the complex $\omega$ plane denoted by $i0$. 

Meanwhile in the $\uparrow$ case, since we account for all possible particle-hole excitations of the Fermi gas, the Green's function is equivalent to the exact solution which can be calculated using a functional determinant approach~\cite{Levitov1993}. To obtain the exact result, one uses the fact that --- for a single impurity --- $\dhd$ and $\dhd \ch_\uparrow$ have the same matrix elements. This enables us to map $\hat{H}_{\med}+\hat{H}_{\rm int}$ in Eq.~\eqref{Eq:Hup2Channel} onto the fermionic, bilinear Hamiltonian
\begin{align} \label{Eq:BiLinear}
    \hat{H}_1 &= \sum_{\kv} (\epsilon_{\kv} - \mu) \fhd_{\kv} \fh_{\kv} + \nu \dhd \dh + g \sum_{\kv} \left( \dhd \fh_{\kv} + \fhd_{\kv} \dh   \right).
\end{align}
Using this, the time-dependent $\uparrow$ impurity Green's function is given by~\cite{Levitov1993}
\begin{align}
    \GM_\uparrow (t) = -i \Theta(t) \det[1 - \hat{n} + \hat{n} e^{i \hat{h}_{\med} t} e^{-i \hat{h}_1 t}], \label{Eq:G_upFDA}
\end{align}
where we have introduced the single-particle counterparts to $\hat{H}_{\med}$ and $\hat{H}_{1}$ denoted $\hat{h}_{\med}$ and $\hat{h}_{1}$, respectively, along with the occupation operator $\hat{n} = 1/[\exponential({\beta \hat{h}_{\med})}+1]$. The $\uparrow$ Green's function in the frequency domain can now be calculated from the Fourier transform of Eq.~\eqref{Eq:G_upFDA}.

\section{Response to a weak Rabi drive}\label{Sect:Applications}
In this section we will explore two applications of our variational approach in the regime of weak Rabi driving: 
 the short-time dynamics of Rabi oscillations and rf spectroscopy beyond linear response. In the case of rf spectroscopy, we observe a striking agreement between our theoretical results and the experimental measurements in Ref.~\cite{Kohstall2012}, which further confirms the accuracy of our approximation in this regime.  

\subsection{Short-time dynamics of Rabi oscillations} \label{Sect:ShortTimeDynamics}

Our relationship between Rabi oscillations and the impurity Green's functions allows us to calculate the short-time dynamics of Rabi oscillations at weak Rabi drive (i.e., $\Omega_0 t \ll 1$). In particular, we remind the reader that we are considering Rabi oscillations initialized in the $\downarrow$ state, where the $\downarrow$ state is noninteracting with the Fermi gas. To calculate these dynamics, it is instructive to first connect the Rabi oscillations to the Ramsey response, which is given by $S(t) = i \mathcal{G}_\uparrow (t)$~\cite{Cetina2016} at $t\geq0$. The relationship between these protocols can be seen perturbatively by taking $\Omega_0 \ll E_F$ %
and expanding the Rabi coupled $\downarrow$ Green's function, Eq.~\eqref{Eq:GCoupledRelation}, in powers of $\Omega_0$:
\begin{align} \label{Eq:FTGreenCoupledPert}
    \GT_\downarrow (\omega) &\simeq G_\downarrow(\omega) + \frac{\Omega_0^2}{4} G_\downarrow(\omega)  G_\uparrow(\omega) G_\downarrow(\omega)+ \mathcal{O}(\Omega_0^4).
\end{align}
Fourier transforming Eq.~\eqref{Eq:FTGreenCoupledPert} and applying Eq.~\eqref{Eq:RabiFromGreen} yields Rabi oscillations (see Appendix \ref{Sect:PertRabi})
\begin{align} \label{Eq:NfromS}
    \mathcal{N}_\downarrow(t)  \simeq   1 - \frac{ \Omega_0^2}{2} \int_0^{t} \dd t' \, (t-t') \Re[  S(t') e^{i \Delta t'}]+ \mathcal{O}(\Omega_0^4).
\end{align}
We point out that Eq.~\eqref{Eq:NfromS} is consistent with the expected result of Fermi's golden rule, since
\begin{subequations}
\begin{align}
    \partial_t \mathcal{N}_\downarrow(t \to \infty ) &\simeq -\frac{\Omega_0^2}2 \int^\infty_0 \dd t' \, \Re[S(t') e^{i \Delta t'}]\label{eq:FermiGoldena}\\
    &\propto A_\uparrow(\Delta), \label{Eq:FermiGolden}
\end{align}
\end{subequations}
where we have introduced the $\uparrow$ spectral function $A_\uparrow(\omega)$, which is related to the Ramsey response via~\cite{mahan2013many}
\begin{align}
    A_\uparrow(\omega) =\frac{1}{\pi} \int_0^\infty \dd t\, \Re[S(t) e^{i \omega t}].
    \label{eq:AfromS}
\end{align}
The limit $t \to \infty$ in Eq.~\eqref{eq:FermiGoldena} should be understood as taking $t\gg \tau_F$ while still requiring $\Omega_0t\ll1$.

We can now calculate the short-time response to a weak Rabi drive. At times %
$t\ll \tau_F$ where only two-body physics is important, the Ramsey response is given by~\cite{Parish2016}
\begin{align}
    S(t) \simeq 1 - \frac{8 e^{-i \pi /4} (m/m_r)^{3/2}}{9 \pi^{3/2}} \left(\frac{t}{\tau_F} \right)^{3/2},
    \label{eq:SR=0}
\end{align}
while for $R^*>0$, it is~\cite{Parish2016}
\begin{align}
    S(t) &\simeq 1 - \frac{(m/m_r)^2}{3 \pi k_F R^*} \left(\frac{t}{\tau_F} \right)^2 \nonumber\\
    &{}\hspace{5em}+ \frac{16 e^{i\pi/4} (m/m_r)^{5/2}}{45 \pi^{3/2} (k_F R^*)^2} \left(\frac{t}{\tau_F} \right)^{5/2}.
\end{align}
Note that these expressions are valid both at zero and at finite temperature provided $t \ll 1/T$ such that thermal effects are irrelevant~\cite{Parish2016}.
Using %
Eq.~\eqref{Eq:NfromS} then yields the early-time response to the Rabi drive
\begin{align} \label{Eq:N(t)R=0}
   \mathcal{N}_\downarrow(t) \simeq 1 -\frac{ \Omega_0^2}{4}t^2 + \frac{8 \sqrt{2}  
   \Omega_0^2  (m/m_r)^{3/2}}{315 \pi ^{3/2}  \tau
   _F^{3/2}} t^{7/2},
\end{align}
for an effective range of $R^*=0$, and
\begin{align} \label{Eq:N(t)R>0}
    \mathcal{N}_\downarrow (t) &\simeq 1 - \frac{ \Omega_0^2} {4} t^2  + \frac{\Omega_0^2}{144}\left(3 \Delta^2+ \frac{2 (m/m_r)^2}{\pi  k_F R^* \tau_F^2} \right) t^4 \nonumber \\ 
       &{} \hspace{5.5em}- \frac{16 \sqrt{2} \Omega_0^2 (m/m_r)^{5/2}}{2835 \pi^{3/2} (k_F R^*)^2 \tau_F^{5/2}} t^{9/2} ,
\end{align}
for $R^*>0$, where $m_r = m m_{\text{im}}/(m + m_{\text{im}})$ with $m_{\text{im}}$ the mass of the impurity. Here we have momentarily reintroduced the impurity mass %
since the short-time dynamics of the Rabi oscillations effectively correspond to a %
transfer out of the %
initial zero momentum $\downarrow$ state, and thus the results hold even for a mobile impurity.

\begin{figure*}
    \centering
    \includegraphics[width=\linewidth]{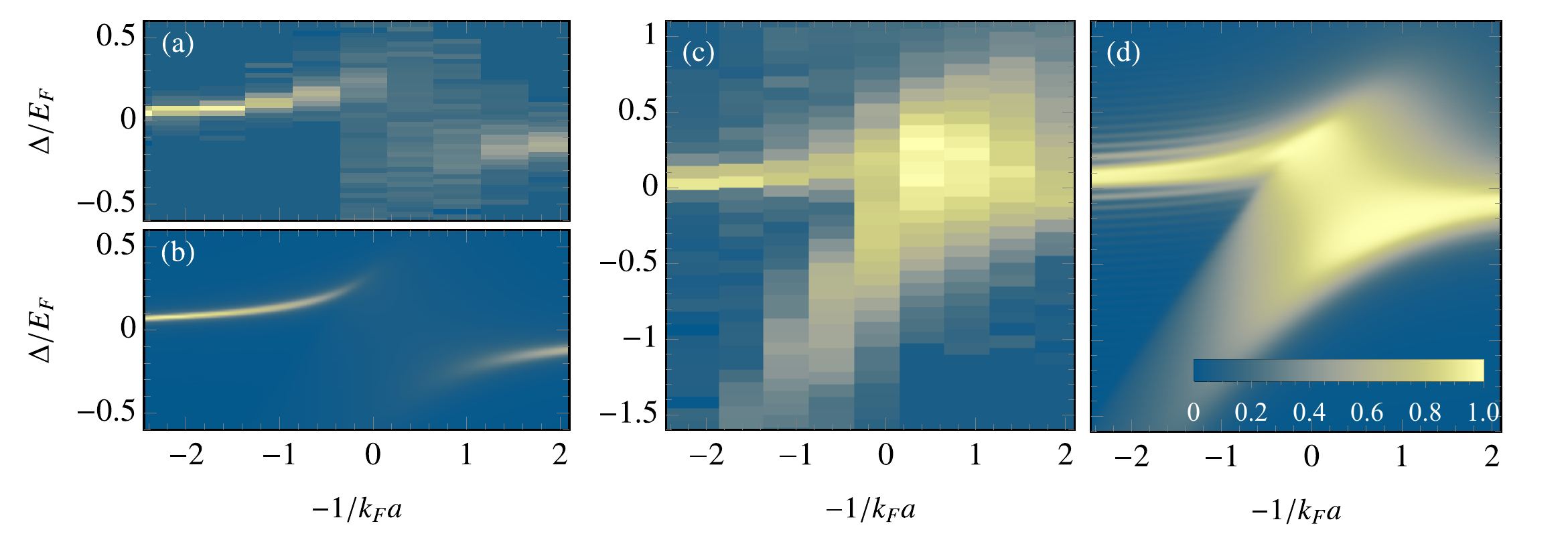}
    \caption{Spectral response of $^{40}$K impurities in a $^{6}$Li Fermi sea as observed in experiment~\cite{Kohstall2012} (a, c) and calculated by our theory (b, d). The density plot shows the fraction of $^{40}$K atoms transferred from the noninteracting $\downarrow$ state into the resonantly interacting $\uparrow$ state [i.e., $\mathcal{N}_\uparrow (\tau_F)$] as a function of detuning $\Delta$ and interaction strength $(k_F a)^{-1}$. To match experiment, panel (a) uses a weak Rabi coupling $\Omega_0/E_F=0.014$, while panel (c) row uses a stronger Rabi coupling $\Omega_0/E_F = 0.14$, which saturates the $^{40}$K absorption and reveals the many-body continuum. The transferred fraction is calculated at time $\Omega_0 \tau_R = \pi$ and $5\pi$ 
    in panels (a,b) and (c,d), respectively. In all panels the temperature is $T/T_F = 0.16$ and the data is normalized to the maximum value of the panel. In the theory panels (b,d) we use an effective range of $k_F R^* = (40/46)^2\times0.95$.}
    \label{fig:KohstallSpectra}
\end{figure*}

In Eqs.~\eqref{Eq:N(t)R=0} and \eqref{Eq:N(t)R>0} we see that the short-time dynamics has a quadratic dependence on time that 
matches the behavior in the absence of any impurity-medium interactions: %
\begin{align}
    \mathcal{N}_\downarrow (t) = 1+\Omega_0^2 \left(-\frac{t^2}{4}+\frac{\Delta ^2
   t^4}{48}+ \mathcal{O} (t^6)\right)+\mathcal{O}(\Omega_0^4).
\end{align}
This is to be expected since both impurity spin states are initially bare particles without any excitations of the Fermi sea.  
However, we observe the emergence of non-analytic behavior in the higher order terms of Eqs.~\eqref{Eq:N(t)R=0} and \eqref{Eq:N(t)R>0}, which is a direct consequence of the renormalization of the contact interactions~\cite{Parish2016}. We also note that the short time dynamics 
of the Rabi oscillations for weak Rabi coupling is independent of the detuning $\Delta$ until terms of order $t^4$ %
(not included in Eq.~\eqref{Eq:N(t)R=0}).

Since the results in Eqs.~\eqref{eq:SR=0}-\eqref{Eq:N(t)R>0} only involve two-body physics (i.e., the two-body impurity-medium T matrix),
%
and they only depend on the medium via the density~\cite{Parish2016}, we furthermore expect that they apply also to the Bose polaron upon a suitable redefinition of $\tau_F$ and $k_F$ in terms of the medium density. Indeed, the short-time behavior of the Ramsey response has already been investigated in a recent experiment~\cite{Skou2021} and agrees with the analytic expression \eqref{eq:SR=0} 
first predicted in Ref.~\cite{Parish2016}.

\subsection{Beyond linear response} \label{Ref:NonLinearSpec}

In experiment, Rabi coupling the impurity spin states represents a standard protocol to determine the spectral response. 
In particular, by virtue of Fermi's golden rule, the transition rate into the $\uparrow$ impurity state at weak Rabi drive is related to the impurity spectral function [see Eq.~\eqref{Eq:FermiGolden}]. This transition rate can also be measured at stronger Rabi drive, in which case one can no longer apply linear response theory. 
On the other hand, as we now show, our variational method %
is capable of simulating this experimental protocol in both the regime of linear response and beyond.

We demonstrate this through simulating the work of Kohstall \textit{et al.} \cite{Kohstall2012}, who employed fermionic $^{40}$K impurity atoms, in two accessible spin states, in a $^{6}$Li Fermi gas at temperature $T/T_F=0.16$. Following the work of Ref.~\cite{Cetina2016}, we take the effective range $k_F R^* \simeq (40/46)^2 \times 0.95$, where the factor of $(40/46)^2$ accounts for 
the assumption of an infinite impurity mass and %
ensures that terms up to order $t^4$ in Eq.~\eqref{Eq:N(t)R>0} are the same in experiment and theory.

Figure~\ref{fig:KohstallSpectra} compares the experimentally measured spectral response of $^{40}$K impurities in a $^{6}$Li Fermi sea~\cite{Kohstall2012} with that calculated within our theory. The simulation closely mirrors the experimental protocol: At a particular interaction strength, detuning, and Rabi drive, we calculate the transferred fraction of impurities after time $\tau_R$ from the $\downarrow$ to the $\uparrow$ state, %
i.e., $\mathcal{N}_\uparrow (t=\tau_R)$. We stress that these parameters are all determined by experiment~\cite{Kohstall2012}, and hence Fig.~\ref{fig:KohstallSpectra} has \textit{no fitting parameters}.

The comparison in Fig.~\ref{fig:KohstallSpectra} is shown for both weak (linear response) and %
stronger (beyond linear response) Rabi drive. By performing the simulation in the time domain, we find that we can accurately capture the experimental spectral response, including the spectral width of the branches. This is despite the absence of impurity recoil in our calculation, which suggests that this does not play a strong role in the spectroscopy at finite temperature, both in the domain of linear response and beyond. We also note that the Blackman pulses~\cite{Kasevich1992} used in experiment, which are time-dependent, are here modelled instead using the time-independent Hamiltonian $\hat{H}_\Omega$, as defined in Eq.~\eqref{Eq:RotWave}. This corresponds to applying a rectangular Rabi pulse and so produces the familiar fringes in the spectral response (panel d); these are apodized by the Blackman envelope in the data (panel c).

The close agreement between our variational Rabi oscillations and the experimental spectra provides strong evidence for the validity of our approximation. It is particularly noteworthy that the comparison holds even at unitarity, where one might expect the co-existence of the $\downarrow$ impurity with multiple particle-hole excitations in the Fermi gas to play an important role. %

\section{Orthogonality catastrophe in Rabi oscillations} \label{Sect:OrthogonalityCatastrophe}
We now investigate the effect of coupling a (quasi)-particle to the Fermi-edge singularity. Since the $\downarrow$ impurity is a bare noninteracting particle while the $\uparrow$ impurity exhibits the orthogonality catastrophe, we achieve this by Rabi coupling the two spin states.

\subsection{Analytical results}
The problem of a fixed impurity in a Fermi sea exhibits the orthogonality catastrophe at zero temperature, where the ground state of the interacting and noninteracting systems are orthogonal~\cite{Anderson1967}. A hallmark of this effect is the formation of a Fermi-edge singularity 
at $\omega_0$ 
in the $\uparrow$ spectral response at low temperature. 
In particular, for $\omega \approx \omega_0$ the spectral response will behave like
\begin{align}
    A_\uparrow(\omega) \sim \Theta(\omega - \omega_0) (\omega-\omega_0)^{\alpha-1},
\end{align}
which we depict in Fig.~\ref{fig:Schematic}(b). Here, the position of this singularity corresponds to
\begin{align}
    \omega_0 = - \int_0^{E_F} \dd E \, \delta(\sqrt{2mE})/\pi,
\end{align}
while the exponent is 
\begin{align} \label{eq:alpha}
    \alpha = \delta(k_F)^2/\pi^2,
\end{align}
where the scattering phase shift is defined via $-k \cot \delta (k)= 1/a+R^*k^2$~\cite{nozieres1969singularities}. Correspondingly, at frequencies $\omega \approx \omega_0$, the $\uparrow$ Green's function is given by~\cite{nozieres1969singularities}
\begin{align} \label{Eq:OrthogGFunc}
    G_\uparrow (\omega) \approx  C (-i)^{1+\alpha} \left[i (\omega_0 - \omega) \right]^{\alpha-1},
\end{align}
where $C$ is a real constant with units of $E_F^{-\alpha}$. Note that here and in the following, it should be understood that $\omega$ contains a small positive imaginary part like in Eq.~\eqref{Eq:GreenDownBare}. 
For positive scattering lengths, the occupation of an impurity-fermion bound state [see Eq.~\eqref{Eq:BoundState}] generates an additional singularity with frequency $\omega_b = \omega_0 - E_F +E_b$ and exponent $\alpha_b = [1+\delta(k_F)/\pi]^2$~\cite{combescot1971infrared}.

\begin{figure*}
    \centering
    \includegraphics[width=0.95\linewidth]{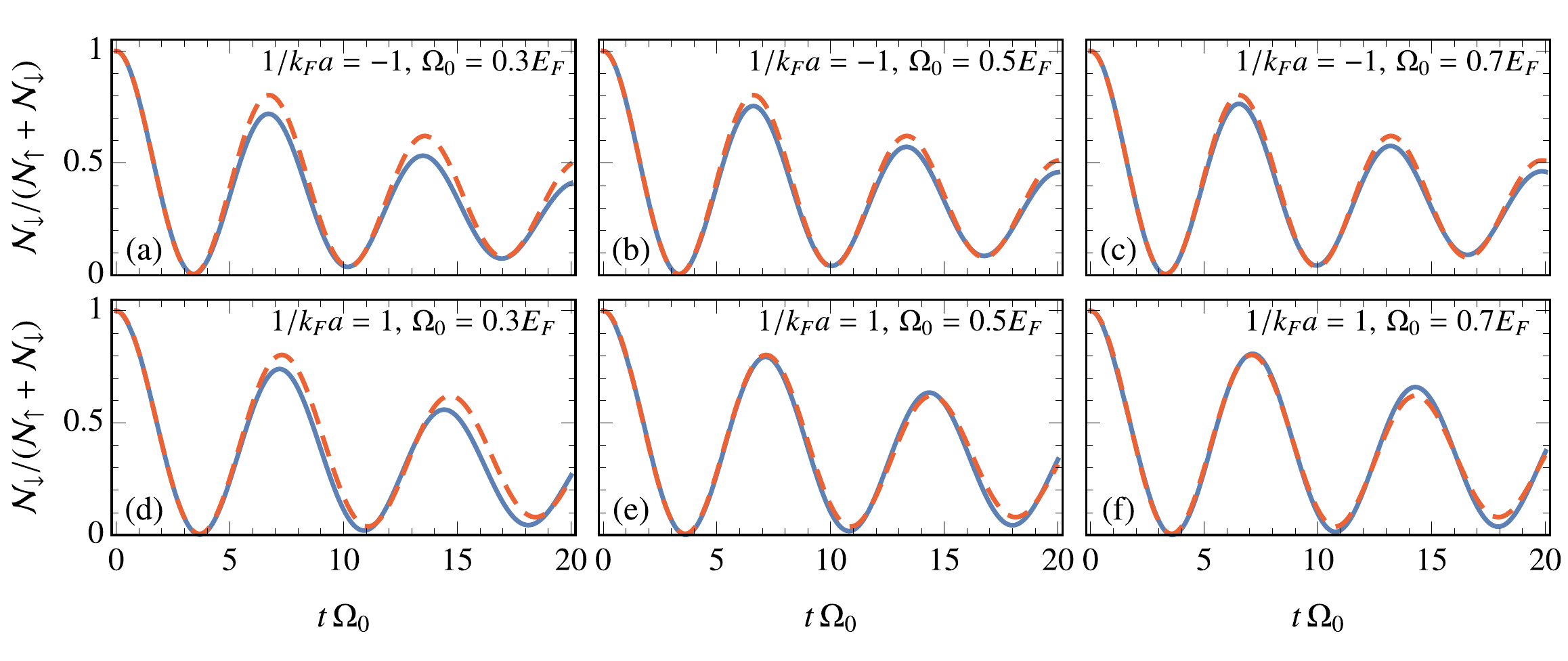}
    \caption{Comparison of the variational (blue) and analytical (orange) Rabi oscillations obtained from Eqs.~\eqref{Eq:RabiFromGreen} and \eqref{Eq:ApproxRabi}, respectively. We calculate these at $R^*=0$ and $T=0.05T_F$, with the detuning set to match the Fermi edge singularity (i.e., $\Delta = \omega_0$). In the top row, $1/k_F a = -1$ and we set $C=0.96 E_F^{-\alpha}$, while in the bottom row, $1/k_F a = 1$ and we set $C=0.83E_F^{-\alpha}$. From left to right in both rows, the Rabi coupling is $\Omega_0/E_F = 0.3$, 0.5 and 0.7.}
    \label{fig:CompExactAndApprox}
\end{figure*}

For a sufficiently weak Rabi drive, the Rabi oscillations are primarily determined by the $\uparrow$ spectral features close to the frequency set by the detuning $\Delta$ from the $\uparrow$-$\downarrow$ transition. 
Thus, when $\Delta = \omega_0$, we can use the approximate form of the Green's function in Eq.~\eqref{Eq:OrthogGFunc} %
and perform an analytical analysis that elucidates the implications of the orthogonality catastrophe for the Rabi oscillations. %
Using Eqs.~\eqref{Eq:GCoupledRelation} and \eqref{Eq:GreenDownBare}, this yields the Rabi-coupled $\downarrow$ Green's function
\begin{align} \label{Eq:ApproxCoupledGreen}
    \GT_\downarrow (\omega) \simeq \frac{1}{\omega - \omega_0 - C (-i)^{1+\alpha} \left[i (\omega_0 - \omega ) \right]^{\alpha-1} \Omega_0^2/4}.
\end{align}

To determine the corresponding Rabi oscillations, %
we require the %
inverse Fourier transform of Eq.~\eqref{Eq:ApproxCoupledGreen}. In Appendix~\ref{Sect:AnalyticalModel} we obtain this using contour integration, which yields Rabi oscillations
\begin{align} \label{Eq:ApproxRabi}
    \mathcal{N}_\downarrow(t) \simeq  \left| \frac{e^{-i \beta_< t} + e^{-i \beta_> t}}{i(2-\alpha)} - I(t) \right|^2,
\end{align}
where
\begin{subequations}
\begin{align}\label{Eq:quasiparticleEnergies}
    \beta_< &= \omega_0 -  \left( \frac{4}{C \Omega_0^2 \, } \right)^ {\frac{1}{\alpha-2}},\\
    \beta_> &=\omega_0+  \left( \frac{4i^{2\alpha}}{C \Omega_0^2 \, } \right)^{\frac{1}{\alpha-2}},
\end{align}
\end{subequations}
and $ I(t)$ is a contour integral arising from the branch cut in $G_\uparrow (\omega)$:
\begin{align} \label{Eq:ContourIntergral}
     I(t) &\equiv e^{- i \omega_0 t } \int_0^{\infty} d \omega\frac{ 2 \omega\, e^{-\omega t}}{\pi }  \nonumber\\
     &\cross\left(\frac{1}{4 \omega
  ^2+e^{\frac{1}{2}i \pi  \alpha } C  \Omega_0^2 \, \omega ^{\alpha
  }}-\frac{1}{4 \omega ^2+e^{-\frac{3}{2} i \pi  \alpha }
  C \Omega_0^2\, \omega ^{\alpha }}\right).
\end{align}
Here, the frequencies $\beta_<$ and $\beta_>$ correspond to the positions of %
emergent quasiparticle peaks in the coupled spectrum $\tilde{A}_\downarrow(\omega)$, as shown schematically in Fig.~\ref{fig:Schematic}.
These two quasiparticles largely determine the behavior of the Rabi oscillations since the branch cut contribution is initially small [$I(t=0)=i\alpha /(\alpha-2)$], and it effectively dampens out before a single period due to the $e^{-t\omega}$ term in Eq.~\eqref{Eq:ContourIntergral}.
Importantly, this implies that the frequency of the Rabi oscillations is approximately given by
\begin{align} \label{Eq:orthogScaling}
    \Omega \simeq \Re[\beta_> - \beta_<] = C_1 E_F (\Omega_0/E_F)^{2/(2-\alpha)},
\end{align}
where $C_1$ is a dimensionless constant~\footnote{The dimensionless constant in the frequency scaling of the Rabi oscillations is given by $C_1 = (4 E_F^{-\alpha}/C)^{1/(\alpha-2)} \{1+\cos[\pi \alpha/(\alpha-2)] \} $.}. Thus, the Rabi oscillation frequency 
scales as a non-trivial power of the bare Rabi coupling, reflecting the Fermi edge singularity. Crucially, the relationship in Eq.~\eqref{Eq:orthogScaling} is  qualitatively different from the case of a mobile impurity, where the frequency is related to the addressed quasiparticle's residue $Z$ and inverse lifetime $\Gamma$ via $\Omega \simeq \sqrt{Z \Omega_0^2 - \Gamma^2}$~\cite{Kohstall2012,Adlong2020} which reduces to a linear scaling $\Omega\simeq \sqrt{Z} \Omega_0$ for small $\Gamma$.

Strictly speaking, the scaling relation in Eq.~\eqref{Eq:orthogScaling} is only exact in the regime $\Omega_0 \ll E_F$ where our approximation is valid, but we also expect it to extend to larger $\Omega_0$ since the leading order corrections mainly affect the fine structure of the oscillations rather than the dominant frequency~\cite{Adlong2020}. Indeed, it has been shown in Ref.~\cite{Knap2013} that the low-energy limit of the Rabi-coupled impurity problem maps onto the spin-boson model with an Ohmic bath~\cite{Leggett87}, where the bare Rabi coupling $\Omega_0$ corresponds to the tunnelling amplitude between the two spin states. In this case, it is well known~\cite{Leggett87} that the renormalized Rabi coupling $\Omega$ has the power-law scaling in Eq.~\eqref{Eq:orthogScaling}, where $E_F$ plays the role of a high-energy cut-off. Thus, our analysis recovers the correct low-energy physics. 

We emphasize that while the power-law scaling of the Rabi frequency has previously been theoretically studied in Ref.~\cite{Knap2013}, the calculations were restricted to zero temperature and the numerics were performed on an effective one-dimensional lattice model. Moreover, the impurity spectral function and the emergent quasiparticles in the coupled system were not considered.

\subsection{Frequency scaling of Rabi oscillations}
We now investigate the power law scaling of %
the Rabi frequency $\Omega$ at experimentally relevant temperatures.  
For simplicity, we focus on the case of $R^*=0$.
To begin our analysis, we first compare the %
analytical approximation in Eq.~\eqref{Eq:ApproxRabi} with the variational Rabi oscillations, Eq.~\eqref{Eq:RabiFromGreen}. This comparison is shown in Fig.~\ref{fig:CompExactAndApprox} for %
$T=0.05T_F$, $1/k_F a = \pm 1$ and for a range of Rabi coupling strengths. In calculating the analytical Rabi oscillations, we use the expression for $\alpha$ in Eq.~\eqref{eq:alpha}, and determine $C$ from fitting the analytical expression to the variational Rabi oscillations at $\Omega_0 = 0.7 E_F$ (for each $1/k_F a$). For $1/k_F a =-1$, we observe that the damping in the analytical approximation is slightly underestimated due to thermal damping effects, while the frequency appears to scale accurately with $\Omega_0$. For $1/k_F a = 1$, the physics is complicated by the additional singularity with frequency $\omega_b$. In particular, we observe slight deviations in the damping and the frequency of the Rabi oscillations with no clear trend. However, in general we find a clear consistency between the simple analytical model and the variational Rabi oscillations.

\begin{figure}
    \centering
    \includegraphics[width=\linewidth]{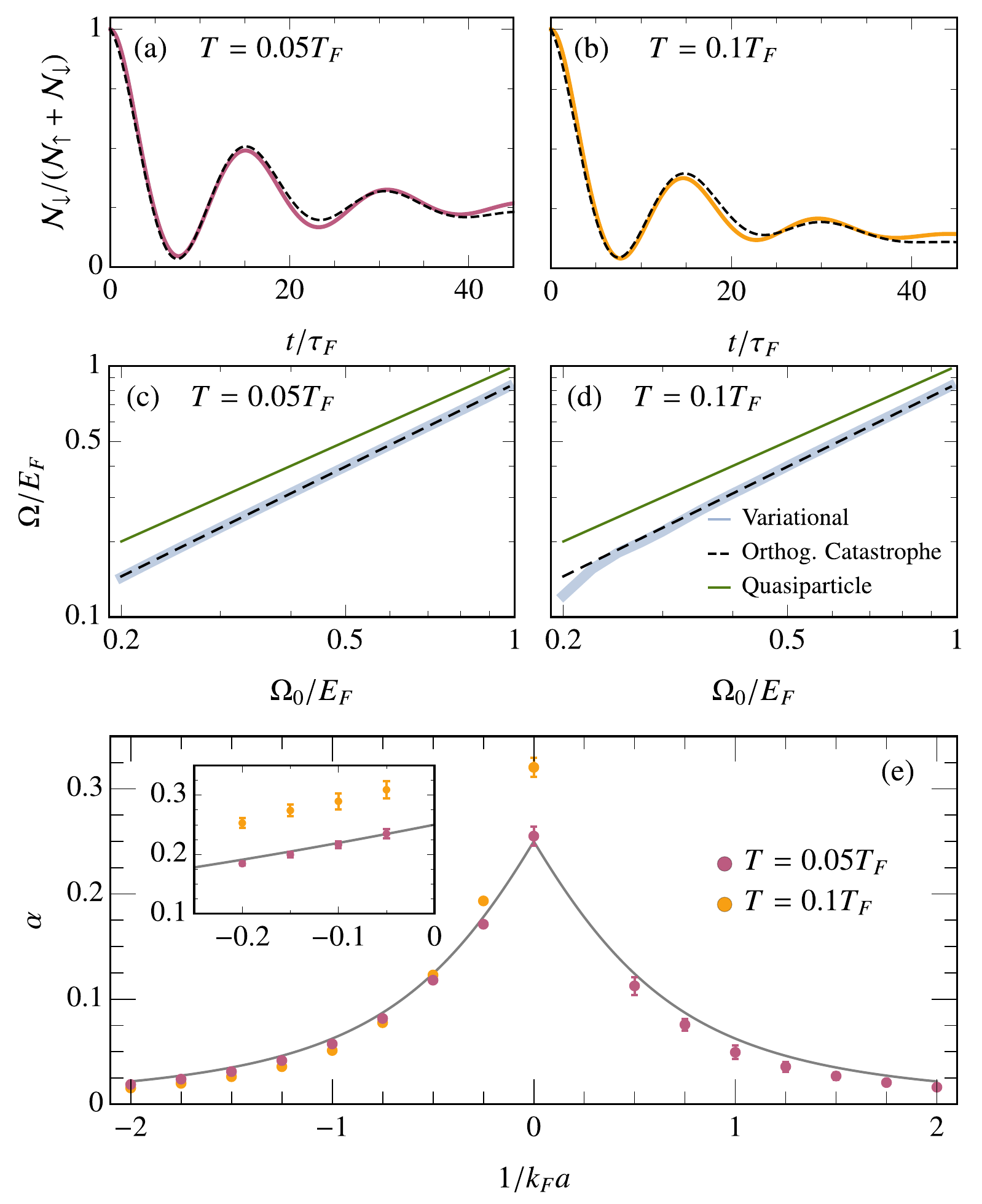}
    \caption{Frequency analysis of the variational Rabi oscillations at $k_F R^* = 0$. In the top row, we show the variational Rabi oscillations (solid line) and their corresponding damped sinusoidal fit (dashed black line) at $1/k_F a = -0.25$, $\Omega_0/E_F=0.5$ and temperature $T/T_F = 0.05$ (a) and $0.1$ (b). In the middle row, we show the frequency of the Rabi oscillations (blue line) as a function of the Rabi drive at $1/k_F a = -0.25$ and $T/T_F = 0.05$ (c) and $0.1$ (d). We compare this against the expected behavior of addressing the orthogonality catastrophe $\Omega = 0.85 E_F\, (\Omega_0/E_F)^{2/(2-\alpha)}$ (dashed black line) and of addressing a quasiparticle of unit residue, $\Omega = \Omega_0%
    $ (green line). %
    In the bottom panel, we plot $\alpha = \delta(k_F)^2/\pi^2$ %
    (gray line) alongside the extracted values of $\alpha$ from fitting the frequency scaling of the Rabi oscillations at $T/T_F = 0.05$ (purple) and $0.1$ (orange). The scaling of the oscillations is fit with $\Omega_0/E_F \in [0.2,0.8]$ (purple, $1/k_F a \leq 0$), $[0.2,0.4]$ (purple, $1/k_F a > 0$) and $[0.4,0.8]$ (orange). Inset: Extracted values of $\alpha$ close to unitarity. We vary the maximum fitting time of the Rabi oscillations from $45 \tau_F$ to $60 \tau_F$, 
    and the spread in results is indicated with error bars.}
    \label{fig:ExtractAlpha}
\end{figure}

In Fig.~\ref{fig:ExtractAlpha} we use the power law scaling of the Rabi frequency %
to simulate an experimentally implementable method for extracting the exponent of the Fermi-edge singularity $\alpha$. %
We expect this method to be experimentally feasible since Rabi oscillations are a well-established experimental protocol~\cite{Kohstall2012,Scazza2017,Oppong2019}. 
In Figs.~\ref{fig:ExtractAlpha}(a) and (b), we show that the variational Rabi oscillations are accurately captured by a damped sinusoid that is typically 
used to extract the frequency of Rabi oscillations from the measured impurity populations~\cite{Scazza2017,Oppong2019}:
\begin{align} \label{Eq:fit}
    \mathcal{N}_\downarrow (t) \simeq b e^{-\Gamma_{\text{bg}}t} + (1-b) e^{-\Gamma_R t} \cos(\Omega t),
\end{align}
where $b$, $\Gamma_{\text{bg}}$, $\Gamma_R$ and $\Omega$ are fitting parameters.
Extracting the frequency $\Omega$ as a function of Rabi drive $\Omega_0$ [panels (c) and (d)], shows two distinct regions of the Rabi oscillations: regions dominated by either thermal broadening or the orthogonality catastrophe.
At finite temperature, the two peaks in the coupled spectrum (i.e., $\beta_<$ and $\beta_>$ in Fig.~\ref{fig:Schematic}) broaden. Consequently, at sufficiently weak Rabi drive where $\Omega_0 \lesssim 2 T$, the two peaks become indistinguishable, which translates into thermally dominated 
dynamics that do not exhibit coherent oscillations.
Once the Rabi coupling is sufficiently large such that it overcomes thermal effects, while still being small enough that it does not completely smear out the power-law part of the spectrum, i.e., for $2T \lesssim \Omega_0 \lesssim E_F$, we find that the orthogonality catastrophe scaling [see Eq.~\eqref{Eq:orthogScaling}] dictates the frequency of the Rabi oscillations. In particular, panels (c) and (d) show that the power-law scaling relation (black dashed lines) accurately matches the slope of the variational Rabi frequency curves (blue solid lines).

In panel (e) of Fig.~\ref{fig:ExtractAlpha} we finally extract $\alpha$ --- the exponent of the Fermi-edge singularity --- through fitting the frequency of the Rabi oscillations in a finite range of $\Omega_0$ values. For negative scattering lengths, this finite range is $\Omega_0 \in [4T, 0.8 E_F]$, such that we are safely in the domain where the physics of the orthogonality catastrophe dominates. 
For positive scattering lengths, the additional singularity at $\omega_b$ limits our ability to probe the singularity at $\omega_0$, which forces us to constrain the fitting range to $\Omega_0 \in [4T, 0.4 E_F]$, and therefore %
we only perform the corresponding analysis %
for $T=0.05 T_F$. As seen in panel (e), we can accurately extract $\alpha$ for all interactions (except those that are strongly repulsive) for $T=0.05 T_F$, and at attractive interactions for $T=0.1T_F$. We have thus demonstrated that the power law scaling of the %
the Rabi oscillation frequency could serve as an experimentally realistic probe of the orthogonality catastrophe.

\section{Conclusion} \label{Sect:Conclusion}
To conclude, we have investigated Rabi oscillations between the $\uparrow$ and $\downarrow$ states of a fixed impurity in a Fermi gas. In approaching this problem, we have introduced a variational method that enables the Rabi oscillations to be approximated from the spin decoupled Green's functions ($G_\uparrow$ and $G_\downarrow$), which can both be calculated exactly. In the regime of weak Rabi drive, we have demonstrated the utility of our theory for two scenarios.
First, we have calculated the exact short-time dynamics of the Rabi oscillations. We find that the dynamics are directly related to the Ramsey response, and we have used this to explore the emergence of non-analytic behavior in the Rabi oscillations. We have argued that these results also hold for a mobile impurity. Second, we have simulated the rf spectroscopy of $^{40}$K impurities in a $^{6}$Li Fermi gas beyond linear response, finding a remarkable agreement with experiment~\cite{Kohstall2012}. To date, the regime beyond linear response %
has not been accurately simulated and thus the agreement with experiment illustrates the power and accuracy of our variational approach. 

Following these applications, we have investigated the effect of coupling a %
quasiparticle to the Fermi-edge singularity and how this manifests in the Rabi oscillations. 
While quasiparticle behavior is absent in the orthogonality catastrophe, we find that the Rabi coupled %
spectrum exhibits two well-defined quasiparticle peaks, where the energy separation of these peaks determines the frequency of the Rabi oscillations. 
This ultimately leads us to show analytically that the frequency of the Rabi oscillations $\Omega$ scales with the Rabi drive $\Omega_0$ as a power law that is determined by the Fermi-edge singularity. We furthermore show that this power-law scaling survives at finite temperature %
in the variational Rabi oscillations. 

In previous studies it has been suggested that the late-time behavior of the Ramsey response could serve as a probe of the orthogonality catastrophe~\cite{Goold2011,Knap2012,Schmidt2018}. From a theoretical perspective, the Ramsey response is an exceptionally clean probe since it measures the $\uparrow$ Green's function in the time domain. However, the Ramsey response is a highly challenging protocol to implement since it requires one to produce an equal superposition of the two non-interacting impurity spin states ($\uparrow$, $\downarrow$) before rapidly %
switching on the interactions in the $\uparrow$ state. 
To date, only one experimental group has realized the Ramsey response using ${}^{40}$K impurities in a $^6$Li gas \cite{Cetina2015,Cetina2016}. On the other hand, Rabi oscillations are a well-established protocol that is relatively simple to implement. Therefore, %
we would expect a frequency analysis of the Rabi oscillations to act as an experimentally feasible probe of the orthogonality catastrophe.

Finally, we point out that in the scenario where the $\downarrow$ state has vanishing mass, while the $\uparrow$ state remains infinitely heavy, 
our calculated variational Rabi oscillations %
become exact for any drive strength. This is due to the infinite recoil energy associated with the $\downarrow$ impurity in the presence of excitations of the Fermi sea. Such a setup could potentially be %
realized %
using a state-dependent potential %
to fix only the $\uparrow$ state. State-dependent lattices have, for instance, been achieved for 
different electronic states of ultracold $^{173}$Yb atoms~\cite{Riegger2018}.

\acknowledgements
We gratefully acknowledge early discussions on Blackman pulses and spectroscopy beyond linear response with Melina Filzinger. We also thank Matteo Zaccanti for sharing the data of Ref.~\cite{Kohstall2012}, and Michael Knap for pointing out the relevant work of Ref.~\cite{Knap2013}.
HSA acknowledges support through an Australian Government Research Training Program Scholarship. 
LDT acknowledges funding from the Australian Research Council Discovery Project LP200100082. 
JL, WEL and MMP acknowledge support from the Australian Research Council Centre of Excellence in Future Low-Energy Electronics Technologies (CE170100039).  JL and MMP are also supported through Australian Research Council Future Fellowships FT160100244 and FT200100619, respectively, and JL acknowledges support from the Australian Research Council Discovery Project DP210101652.

\appendix

\onecolumngrid

\newcommand{\balpha}{\boldsymbol{\alpha}}
\newcommand{\mH}{\mathcal{H}}
\newcommand{\up}{\uparrow}
\newcommand{\down}{\downarrow}

\section{Variational approach to the spin coupled Green's function} \label{App:DetailsGreenFunc}
Here we provide details on the matrix structure of the spin coupled Green's function, as given in Eq.~\eqref{Eq:GCoupledRelation}. In particular, we will show that this matrix structure is exact within the variational operator operator basis $\{ \hat{O}_j \} = \{\ch_\downarrow , \ch_\uparrow, \fhd_{\qv} \fh_{\kv} \ch_{\uparrow}, \dots \}$. As discussed in Sec.~\ref{Sect:VarRabiOsc}, the expansion coefficients $\{\alpha^\sigma_j \}$ can be calculated from minimizing the `error quantity'
\begin{align}
    \mathcal{E}(t) \equiv \Tr[\rhoh_0\hat\epsilon (t)\hat\epsilon^\dagger (t)],
    \label{eq:error}
\end{align}
where the operator $\hat\epsilon (t) \equiv i\partial_t \ch(t) - \comm{\ch(t)}{\Hat{H}}$. Using the minimization condition $\partial \mathcal{E}(t)/\partial  \dot{\alpha}^\sigma{}^*(t)=0$ and taking the stationary condition $\alpha_j^\sigma = \alpha^\sigma_j (0) e^{-i \omega t} \equiv \alpha^\sigma_j e^{-i \omega t}$, one arrives at a set of linear equations of the form $\mathcal{H}\balpha=\omega\balpha$:
\begin{align} \label{Eq:SetOfEq}
\renewcommand{\arraystretch}{1.5}
\left(\begin{array}{@{}c|c@{}}
  \mathcal{H}_\downarrow &
  \begin{matrix}
  \frac{\Omega_0}{2} & 0 & \dotsm & 0
  \end{matrix}
\\ \hline
  \begin{matrix}
  \frac{\Omega_0}{2} \\ 0 \\ \vdots \\ 0 \end{matrix}
  & 
   \mathcal{H}_\uparrow
\end{array}\right) \mqty(\alpha^\downarrow_0 \\ \alpha_0^\uparrow \\ \alpha^\uparrow_{\kv} \\ \vdots)=    \omega \mqty(\alpha^\downarrow_0 \\ \alpha_0^\uparrow \\ \alpha^\uparrow_{\kv} \\ \vdots).
\end{align}
Here $\mathcal{H}_\downarrow=\Delta$ and $\mathcal{H}_\uparrow$ are both independent of $\Omega_0$ and constitute the corresponding  matrices for the two spin states in the absence of Rabi coupling.
We emphasize that the temperature of the medium is incorporated into $\mathcal{H}$ in Eq.~\eqref{Eq:SetOfEq} since the error minimization in Eq.~\eqref{eq:error} explicitly includes temperature. Formally, solving the system of equations yields the eigenvalues $E_n$ and corresponding eigenvectors $\balpha^{(n)}=(\alpha_0^{\downarrow(n)},\alpha_0^{\uparrow(n)},\alpha_{\kv}^{\uparrow(n)},\dots)$.

To proceed, we consider the relation between $\alpha_0^\downarrow(t)$ and the variational coefficients~\cite{Liu2019}
\begin{align}
    \alpha_0^{\downarrow}(t)=\sum_n |\alpha_0^{\downarrow(n)}|^2e^{-iE_nt}.
\end{align}
This allows us to obtain the impurity Green's function from Eq.~\eqref{eq:Gdownt}:
\begin{align}
    \Tilde{\GM}_\downarrow(t) = -i \Theta(t)\sum_n |\alpha_0^{\downarrow(n)}|^2e^{-iE_nt},
\end{align}
with Fourier transform
\begin{align}
\tilde G_\downarrow(\omega)=\sum_n\frac{|\alpha_0^{\downarrow(n)}|^2}{\omega-E_n+i0}.
\end{align}
Representing the matrix $\mathcal{H}=\sum_n {\boldsymbol{\alpha}^{(n)}} E_n{\boldsymbol{\alpha}^{(n)}}^\dag$, we find the Green's function in terms of $\mathcal{H}$
\begin{align}
\tilde G_\downarrow(\omega)=\left(\frac1{\omega-\mathcal{H}+i0}\right)_{11}.
\end{align}

The Green's function can now be calculated by inverting $\omega-\mathcal{H}$:
\begin{align} \label{Eq:MatrixHam}
\renewcommand{\arraystretch}{1.5} 
\left(\begin{array}{@{}c|c@{}}
  \omega- \mH_\down &
  \begin{matrix}
  \frac{\Omega_0}{2} & 0 & \dotsm & 0
  \end{matrix}
\\ \hline
  \begin{matrix}
  \frac{\Omega_0}{2} \\ 0 \\ \vdots \\ 0 \end{matrix}
  & 
  \omega - \mathcal{H}_\up
\end{array}\right) \equiv \left(\begin{array}{@{}c|c@{}}
  A & B^T \\ \hline
  B & C
\end{array}\right) .
\end{align}
In particular, the inverse of Eq.~\eqref{Eq:MatrixHam} can be analytically calculated through blockwise inversion~\cite{Bernstein2009}
\begin{align} \label{Eq:InverseSymbols}
\renewcommand{\arraystretch}{2} 
\left(\begin{array}{@{}c|c@{}}
  A & B^T \\ \hline
  B & C
\end{array}\right)^{-1} =\frac{1}{\eta} \left(\begin{array}{@{}c c@{}}
   1 & - B^T C^{-1}\\
    -C^{-1}B & \eta C^{-1}+ C^{-1} B B^T C^{-1}
\end{array}\right),
\end{align}
where
\begin{align}
    \eta = A - B^T C^{-1} B = G_{\uparrow}(\omega) \left[ G^{-1}_{\uparrow}(\omega) G^{-1}_{\downarrow}(\omega) - \frac{\Omega_0^2}{4} \right].
\end{align}
Here we have the used the fact that
\begin{align}
    B^T C^{-1} B = \frac{\Omega_0^2}{4}\left(\frac1{\omega-\mH_\up+i0}\right)_{11}=\frac{\Omega_0^2}{4}\sum_n \frac{|\bar\alpha^\uparrow_{0}{}^{(n)}|^2}{\omega - \bar E_n+i0} 
    =\frac{\Omega_0^2}{4} G_\uparrow(\omega),
\end{align}
where the bar indicates coefficients and energies calculated in the absence of Rabi coupling. Since the variational approach becomes exact in the limit of an infinite number of excitations of the medium, the $\uparrow$ Green's function is exact and can be calculated using the functional determinant approach.

The spin-coupled Green's function is then the $2\times 2$ matrix in the top left hand corner of Eq.~\eqref{Eq:InverseSymbols}, i.e.,
\begin{align} 
\renewcommand{\arraystretch}{2} 
 \GT(\omega)&=\frac{1}{G^{-1}_{\uparrow}(\omega) G^{-1}_{\downarrow}(\omega) - \frac{\Omega_0^2}{4}} \left(\begin{array}{@{}c c@{}}
   G^{-1}_\uparrow(\omega) & - \frac{\Omega_0}{2}\\
    - \frac{\Omega_0}{2} & G^{-1}_\downarrow(\omega)
\end{array}\right)\\
&= \mqty( G^{-1}_\downarrow(\omega) & \frac{\Omega_0}{2} \\
\frac{\Omega_0}{2} &  G^{-1}_\uparrow(\omega))^{-1},\label{Eq:2x2Structure}
\end{align}
which matches Eq.~\eqref{Eq:GCoupledRelation}.

\section{Perturbative analysis of Rabi oscillations}
\label{Sect:PertRabi}
In Sec.~\ref{Sect:ShortTimeDynamics}, we presented a perturbative connection between Rabi oscillations and the Ramsey response, which we show the details of here. We begin by considering Eq.~\eqref{Eq:FTGreenCoupledPert} which states that for small Rabi coupling $\Omega_0$, the $\downarrow$ coupled Green's function can be expanded as
\begin{align} \label{Eq:FTGreenCoupledPertAppendix}
    \GT_\downarrow (\omega) &\simeq \frac{1}{\omega - \Delta +i0 } + \frac{1}{4} \frac{G_\uparrow(\omega)}{(\omega - \Delta+i0)^2} \Omega_0^2 + \mathcal{O}(\Omega_0^4).
\end{align}
The corresponding retarded Green's function is
\begin{subequations}
\begin{align}
    \tilde{\mathcal{G}}_\downarrow(t) &\simeq-i \Theta(t) e^{-i \Delta t} + \frac{\Omega_0^2}{8 \pi} \int \dd \omega \frac{G_\uparrow(\omega+i0)}{(\omega - \Delta+i0 )^2} e^{-i \omega t}  + \mathcal{O}(\Omega_0^4) \label{Eq:LineA}\\
     &= -i \Theta(t) e^{-i \Delta t} - \frac{\Omega_0^2}{8 \pi} \int \dd \omega \, \dd t_1\,  \dd t_2\, \mathcal{G}_\uparrow(t_1) e^{i \omega t_1} \Theta(t_2) t_2 e^{-i \Delta t_2} e^{i \omega t_2} e^{-i \omega t} + \mathcal{O}(\Omega_0^4) \label{Eq:LineB}\\
      &=  -i \Theta(t) e^{-i \Delta t} - e^{-i \Delta t} \Theta(t)  \frac{\Omega_0^2}{4} \int_0^{t} \dd t' \, (t-t') \mathcal{G}_\uparrow(t') e^{i \Delta t'} + \mathcal{O}(\Omega_0^4).
\end{align}
\end{subequations}
In moving from line \eqref{Eq:LineA} to \eqref{Eq:LineB}, we have used the fact that
\begin{align}
    \frac{1}{(\omega - \Delta  + i0)^2} = - \int \dd t \,e^{i \omega t} \Theta(t) t e^{-i \Delta t} .
\end{align}
Given that $\mathcal{G}_\uparrow(t) = -i \Theta (t) S(t)$, the Rabi oscillations, for times $t \geq0$, are given by
\begin{align} 
    \mathcal{N}_\downarrow(t) &\simeq | \tilde{\mathcal{G}}_\downarrow(t)|^2\nonumber \\
    &=   1 - \frac{ \Omega_0^2}{2} \int_0^{t} \dd t' \, (t-t') \Re[  S(t') e^{i \Delta t'}]+ \mathcal{O}(\Omega_0^4). \label{Eq:NfromSSup}
\end{align}
We point out that Eq.~\eqref{Eq:NfromSSup} is consistent with Fermi's golden rule, which can be seen by evaluating
\begin{align}
    \lim_{t\to \infty}\mathcal{N}'_\downarrow(t) &\simeq - \frac{\Omega_0^2}{2} \int_0^\infty \dd t' \Re[S(t') e^{i \Delta t'}] + \mathcal{O}(\Omega_0^4)\nonumber \\
    &= -  \frac{\Omega_0^2}{2} \pi A_\uparrow(\Delta) +\mathcal{O}(\Omega_0^4), \label{Eq:FermiGoldenSup}
\end{align}
where the prime denotes a derivative in time $t$ and we have used Eq.~\eqref{eq:AfromS}.
The connection between the Rabi oscillations and the Ramsey response is made more explicit by noting
\begin{align} \label{Eq:DDNtoS}
    \mathcal{N}_\downarrow''(t) &\simeq -\frac{ \Omega_0^2}{2}\Re[  S(t) e^{ i \Delta t}] + \mathcal{O}(\Omega_0^4).
\end{align}

\section{Analytical model of the orthogonality catastrophe in Rabi oscillations} \label{Sect:AnalyticalModel}

\begin{figure}
    \centering
\begin{tikzpicture}[decoration={markings,
mark=at position 2.356cm with {\arrow[line width=1pt]{<}},
mark=at position 5.5 cm with {\arrow[line width=1pt]{<}},
mark=at position 8.6 cm with {\arrow[line width=1pt]{<}}
,
mark=at position 12.9cm with {\arrow[line width=1pt]{<}}
,
mark=at position 18.26 cm with {\arrow[line width=1pt]{<}}
}
]
\def\gap{0.2}
\def\bigradius{3}
\def\littleradius{0.5}

\draw[help lines,->] (-1.15*3,0) -- (1.15*3,0) coordinate (xaxis);
\draw[help lines,->] (0,-3*1.15) -- (0,0.9) coordinate (yaxis);

\draw[black, thick,    postaction={decorate}]  
(-1*\bigradius, 0.1) arc (180:260: 1*\bigradius) -- (-0.520945,-0.05)  (-0.520945+0.04*\bigradius,-0.05)  arc (0:180 : 0.02*\bigradius)  (-0.520945+0.04*\bigradius,-0.05) --  (-0.520945+0.04*\bigradius,-2.97309+0.1)  arc (262.32:360: 1*\bigradius) -- (-1*\bigradius, 0.1);

\draw[help lines]
(-0.520945+0.06,-0.1)--(-0.520945+0.06,0.1);

\draw[fill] (-1,0) circle (1pt) node [pin={below left:$\beta_<$}] {};

\draw[fill] (1,-0.25) circle (1pt) node [pin={below right:$\beta_>$}] {};

\node[below] at (xaxis) {$\Re \omega $};
\node[left] at (yaxis) {$\Im \omega $};
\node at (2.3,-2.3) {$C_1$};
\node at (-0.520945+0.025,0.3) {$\omega_0$};
\node at (-0.2,-2) {$C_{2}$};

\end{tikzpicture}
\caption{Contours for the inverse Fourier transform of $\GT(\omega)$ in Eq.~\eqref{Eq:InvFT}. Considered as function of complex $\omega$, $\GT(\omega)$ has a branch cut along $\omega = \omega_0 - i x$ for $x \in [0, \infty)$ and has simple poles at $\beta_<$ and $\beta_>$.}
\label{Fig:ContourIntegral}
\end{figure}
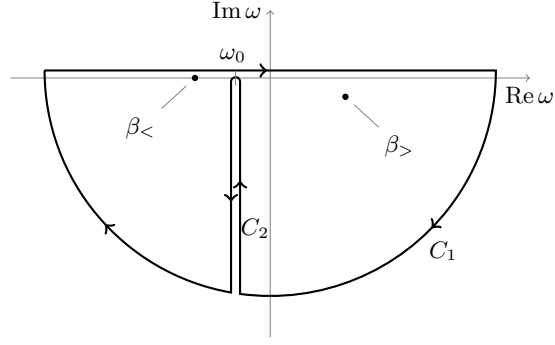

We now present the derivation that led to Eq.~\eqref{Eq:ApproxRabi} in the main text, which allowed us to demonstrate signatures of the orthogonality catastrophe in the Rabi oscillations. As given in Eq.~\eqref{Eq:ApproxCoupledGreen}, we use the approximate Rabi coupled $\downarrow$ Green's function %
\begin{align} \label{Eq:ApproxCoupledGreenAppendix}
    \GT_\downarrow (\omega) = \frac{1}{\omega  - \omega_0 - C (-i)^{1+\alpha} \left[i (\omega_0 - \omega ) \right]^{\alpha-1} \Omega_0^2/4}.
\end{align}
To calculate the corresponding Rabi oscillations, we require the inverse Fourier transform of Eq.~\eqref{Eq:ApproxCoupledGreenAppendix}
\begin{align} \label{Eq:InvFT}
    \Tilde{\GM}_\downarrow (t) =  \int \frac{d \omega}{2\pi} \, e^{-i \omega t} \Tilde{G}_\downarrow (\omega+i0).
\end{align}
In order to perform this integral, we must examine the behavior of $\GT_\downarrow (\omega)$ in the complex plane. First, we observe that the use of the $\uparrow$ Green's function $G_\uparrow (\omega)$ containing the power law singularity, Eq.~\eqref{Eq:OrthogGFunc}, induces a branch cut in $\GT_\downarrow (\omega)$, which we take to be at $\omega = \omega_0 - i x$ for $x \in [0, \infty)$ (see Fig.~\ref{Fig:ContourIntegral}). This branch cut is equivalent to taking the principal value of the complex logarithm to be $\text{Log} (\omega) = \ln |\omega| + i \Arg(\omega)$, where $\Arg(\omega) \in (- \pi, \pi)$ is the principal complex argument. Second, $\GT_\downarrow (\omega)$ has two simple poles, as we now show.

To simplify the notation when deriving the poles, we introduce $\Cb \equiv C \Omega_0^2/4$ and $\omegab \equiv i(\omega_0 - \omega)$. The equation we need to solve is
\begin{align} \label{Eq:omegabEq}
    \omegab \left[ 1 - \Cb (-i)^{2+\alpha} \omegab^{\alpha-2} \right]=0.
\end{align}
In the cases of interest, we have $0 \leq \alpha \leq 1/4$, which shows that the apparent pole at $\omegab=0$ must be excluded (i.e., the component in the bracket diverges like $\omegab^{\alpha-2}$ close to zero). The remaining possible solutions take the form
\begin{subequations}
\begin{align}
    \omegab_n &= \left( \frac{1}{\Cb} \right)^{\frac{1}{\alpha-2}} \exp[\frac{i \Arg \left( i^{2+\alpha} \right) + 2\pi i n}{\alpha-2}]\\
    &= \left( \frac{1}{\Cb} \right)^{\frac{1}{\alpha-2}} \exp[\frac{i \pi (\alpha + 2 + 4n)}{2(\alpha-2)}], \label{Eq:omegabPossible}
\end{align}
\end{subequations}
where $n \in \mathbb{Z}$. In order to be a solution on the principal branch, the argument of Eq.~\eqref{Eq:omegabPossible} must lie between $-\pi$ and $\pi$, which is only satisfied in the case of $n=-1$ and $n=0$. Thus, we find that $\GT_\downarrow ( \omega)$ has two poles:
\begin{align}
    \beta_< &= \omega_0 -  \left( \frac{4}{C \Omega_0^2 \, } \right)^ {\frac{1}{\alpha-2}},\hspace{1cm}
    \beta_> =\omega_0+  \left( \frac{4i^{2\alpha}}{C \Omega_0^2 \, } \right)^{\frac{1}{\alpha-2}},
\end{align}
where $\beta_< \in \mathbb{R}$ and $\beta_< < \omega_0$ while $\beta_> \in \mathbb{C}$ and $\Re \beta_> > \omega_0$. Finally, both of the poles are simple and their residue can be calculated through applying L'H\^{o}pital's rule:
\begin{align}
    \text{Res}(\GT_\downarrow, \beta_i) &= \frac{1}{1-(\alpha-1)C (-i)^{2+\alpha} [i(\omega_0-\beta_i)]^{\alpha-2} \Omega_0^2/4}
    = \frac{1}{2-\alpha}, \label{Eq:CalculatedResidue}
\end{align}
where $i \in \{<,>\}$. We have checked numerically that no further poles exist.

We now use the contours shown in Fig.~\ref{Fig:ContourIntegral} to perform the inverse Fourier transform. Here, the contribution from the end-point of the branch cut 
at $\omega = \omega_0$ vanishes, and %
likewise the contribution from the large arc $C_1$ is zero since $\GT_\downarrow (\omega)$ decays like $1/\omega$ for large $\omega$. We thus find that
\begin{align}
    \frac{1}{2\pi} \int_{-\infty}^{\infty} \dd \omega \,e^{-i \omega t} \GT_\downarrow (\omega+i0)  = -\frac{1}{2 \pi} \left[ 2\pi i \left( \frac{e^{-i \beta_< t } + e^{-i \beta_> t }}{2-\alpha} \right) + \int_{C_2} \dd \omega \, e^{-i \omega t} \GT_\downarrow (\omega)  \right].
\end{align}
Here the integral arising from the branch cut in $\GT_\downarrow(\omega)$ is given by
\begin{subequations}
\begin{align}
  \frac{1}{2\pi }\int_{C_2} \dd \omega \,e^{-i \omega t} \GT_\downarrow (\omega)  &=  \lim_{\epsilon \to 0^+} \frac{1}{2\pi } \left(\int_{\omega_0 + \epsilon - i \infty}^{\omega_0 + \epsilon} \dd \omega \,e^{-i \omega t} \GT_\downarrow (\omega)  +\int_{\omega_0 - \epsilon}^{\omega_0 - \epsilon - i \infty} \dd \omega \, e^{-i \omega t} \GT_\downarrow (\omega)  \right)\\
  &= e^{- i \omega_0 t }\int_0^{\infty} \dd \omega  \frac{2 \omega \, e^{- \omega t }  }{\pi } \left(\frac{1}{4 \omega
  ^2+e^{\frac{1}{2}i \pi  \alpha } C  \Omega_0^2 \, \omega ^{\alpha
  }}-\frac{1}{4 \omega ^2+e^{-\frac{3}{2} i \pi  \alpha }
  C \Omega_0^2\, \omega ^{\alpha }}\right)\\
  &\equiv I( t).
\end{align}
\end{subequations}

As discussed in the main text, the corresponding Rabi oscillations are
\begin{align}
    \mathcal{N}_\downarrow (t) &\simeq |\tilde{\mathcal{G}}_\downarrow (t)|^2
    = \left| \frac{e^{-i \beta_< t} + e^{-i \beta_> t}}{i(2-\alpha)} - I(t) \right|^2.
\end{align}
We point out that in the limit of weak interactions (i.e., $\alpha \to 0$) the Rabi oscillations reduce to
\begin{align}
    \mathcal{N}_\downarrow (t) &= \cos(\frac{\sqrt{C} \Omega_0}{2}t)^2,
\end{align}
which matches the known result of Rabi oscillations between a discrete state and an infinitely long-lived quasiparticle with residue $C$~\cite{Kohstall2012,Adlong2020}.

\twocolumngrid

\bibliography{polaron_bibliography}

\end{document}